\documentclass[12pt,preprint]{aastex}

\newcommand{\HeI}{\ion{He}{1}~}
\newcommand{\metT}{\mbox{$2~^{3}S~$}}

\slugcomment{To be published in Astrophysical Journal}

\shorttitle{Radiative Transfer of He I}
\shortauthors{Benjamin, Skillman, \& Smits}

\begin{document}


\newlength{\bigpicsize}
\setlength{\bigpicsize}{4.5in}
\newlength{\smpicsize}
\setlength{\smpicsize}{3.5in}

\title{Radiative Transfer Effects in He I Emission Lines}


\author{Robert A. Benjamin}
\affil{University of Wisconsin-Madison, Dept. of Physics, 1150 University
Avenue, Madison, WI 53706; benjamin@wisp.physics.wisc.edu}

\author{Evan D. Skillman}
\affil{Department of Astronomy, University of Minnesota, Minneapolis, MN
55455;
skillman@astro.umn.edu}

\author{Derck P. Smits}
\affil{Dept Math \& Astronomy, University of South Africa, PO Box 392,
Pretoria
0003 South Africa; smitsdp@unisa.ac.za}

\begin{abstract}

We consider the effect of optical depth of the \metT level on the
nebular recombination spectrum of He I for a spherically symmetric
nebula with no systematic velocity gradients. These calculations,
using many improvements in atomic data, can be used in place of the
earlier calculations of Robbins. We give representative Case B line
fluxes for UV, optical, and IR emission lines over a range of physical
conditions: $T=5000-20000$ K, $n_{e}=1-10^{8}~ {\rm cm^{-3}~}$, and
$\tau_{3889}=0-100$. A FORTRAN program for calculating emissivities for
all lines arising from quantum levels with $n \leq 10$ is also available
from the authors.

We present a special set of fitting formulae for the physical
conditions relevant to low metallicity extragalactic H II regions:
$T=12,000-20,000$ K, $n_{e}=1-300~ {\rm cm^{-3}~}$, and $\tau_{3889}<$
2.0).  For this range of physical conditions, the Case B line fluxes
of the bright optical lines 4471 \AA, 5876 \AA, and 6678 \AA, are
changed less than 1\%, in agreement with previous studies.  However,
the 7065 \AA\ corrections are much smaller than those calculated by
Izotov \& Thuan based on the earlier calculations by Robbins.  This
means that the 7065 \AA\ line is a better density diagnostic than
previously thought.  Two corrections to the fitting functions
calculated in our previous work are also given.

\end{abstract}



\section{Introduction}

In a previous paper (Benjamin, Skillman, \& Smits 1998, Paper I), we
presented a calculation of the nebular recombination line spectrum of
\HeI which combined the atomic data and recombination cascade of Smits
(1991; 1996) with the collision strengths data of Sawey \& Berrington
(1993). The resulting Case B line fluxes were tabulated, fitting
functions were provided, and estimates of uncertainties in atomic data
on the helium abundance estimates were also given.

An additional uncertainty in estimating helium abundances is the
effect of radiative transfer. By assuming Case B, we have assumed that
most nebulae have a large optical depth in transitions arising from
the 1 $^{1}S$ level.  However, a significant population of electrons
may also build up in the metastable \metT level. Absorption and
reprocessing of photons from the \metT level to upper levels will
alter the recombination cascade and resulting helium abundance
estimates.

The last full treatment of the recombination cascade combined with
optical depth was provided by Robbins (1968), and this work remains the
classic reference for most researchers. Additional work on this
problem has been done by Almog \& Netzer (1989), Proga, Mikolajewska,
\& Kenyon (1994), and Sasselov \& Goldwirth (1995). However, these
papers, while they considered the radiative transfer problem, used
models of the helium atomic physics with an accuracy less than the
1\% level necessary for meaningful cosmological constraints. As a
result, while they were useful in demonstrating the
{\it relative} effect of radiative transfer; the
{\it absolute} effect has not been addressed since the
improvements in the atomic data. We address this problem here.

This work is of particular interest to the derivation of the
primordial helium abundance.  Accuracy in the atomic
physics of better than 1\% is required to produce abundance estimates
that provide meaningful constraints on models of Big Bang
nucleosynthesis.  Currently, observational studies are claiming
accuracies on order of less than 1\% (e.g., Y$_p$ $=$ 0.2452 $\pm$ 0.0015, 
Izotov et al.\ 1999; Y$_p$ $=$ 0.2345 $\pm$ 0.0026, Peimbert et al.\ 2000).
In Paper I we performed a Monte Carlo exercise to estimate the uncertainty
in the theoretical emissivities due to uncertainties in the input atomic
data and obtained an estimate of 1.3 to 1.5\%.
Recently, Olive \& Skillman (2001) have demonstrated that
minimization routines used to solve for physical conditions and He
abundances simultaneously can result in erroneous solutions due
to the degeneracies of the dependencies of the helium emission lines
on the various physical conditions (i.e., temperature, density,
optical depth, and underlying absorption).  Obviously, using accurate atomic
physics in these calculations reduces the systematic uncertainties
in the abundance calculations.

\section{Calculations}

For this work, we consider the radiative transfer of a spherically
symmetric nebula with no systematic expansion. We assume that the
recombination is in the case B limit, so that the only radiative
transfer effects are those that arise from the metastable \metT
level. The assumption of no systematic expansion is justified
by detailed studies of the velocity fields of giant extragalactic
HII regions (e.g., Yang et al.\ 1996, and references therein) which
show them to be dominated by turbulent, and not organized, motions.
We parameterize our calculations in terms of the line center
optical depth of the 3889 \AA\ line,
$\tau_{3889}=n(\metT)\sigma_{3889}R_{s}$, where $n(\metT)$ is the
density of atoms in the metastable state.\footnote{Note that in the text 
we refer to emission lines using their wavelength in air. However, 
the tables use the vacuum wavelengths derived from the energy levels
described in Paper I.} The cross section at line
center is given by

\begin{equation}
\sigma=(6.73 \times 10^{-31}{\rm cm^{-2}}) ~A \lambda_{\AA}^{3}
        \left[166\frac{T_{4}}{m_{He}}+\xi^{2}\right]^{-1/2}~,
\end{equation}

\noindent where $A({\rm s^{-1}})$ is the spontaneous transition probability;
$\lambda_{\AA}$ is the wavelength in Angstroms; the gas temperature is
$T_{4}=T/(10^{4}~{\rm K})$; $m_{He}=4.0$, and the ``turbulent''
velocity, $\xi$, is typically $40~{\rm km~s^{-1}}$.

For a nebula of Stromgren radius, $R_{s}=(30~{\rm pc})
Q_{48}^{1/3}n_{H}^{-2/3}$, the optical depth is

\begin{equation}
\tau=9.68 \times 10^{-14}
Q_{48}^{1/3}n_{H}^{1/3}p_{-7}A_{6}T_{4}^{-1/2}\lambda^{3}_{\AA}~,
\end{equation}

\noindent where $Q_{48}=Q_{H}/(10^{48}~{\rm cm^{-3}})$ is the luminosity of
hydrogen ionizing photons, $n_{H}$ (${\rm cm^{-3}}$) is the hydrogen
particle density, $A_{6}=A/(10^{6}~{\rm s^{-1}})$ is the spontaneous
transition probability, and $p_{-7}=(n(\metT)/n_{He+})/10^{-7}$ is the
population in the metastable \metT level. A helium abundance of 10\%
by number is assumed.  Turbulent velocities will lower this optical
depth, and can be accounted for by replacing $T_{4}$ with
$T_{4}+m_{He} \xi^{2}/166$, where $\xi$ is in ${\rm km~s^{-1}}$.  For
low densities, the population in the \metT level is in the range of
$p_{-7}=$ 1-4 (See Table 8 in Paper I.)  For fixed $T$ and $p$, the
variation of optical depth with ionizing luminosity and particle
density is shown in Figure 1.

\begin{figure}[ht!]
\includegraphics[angle=0,totalheight=\smpicsize]{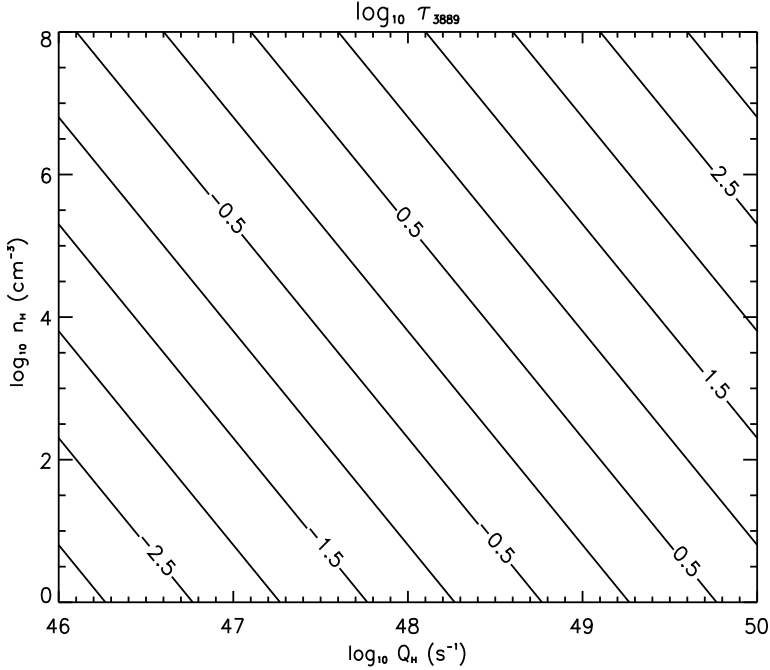}
\caption{ Dependence of optical depth in 3889 \AA\  line center as a
function of ionizing luminosity of source, $Q_{H}$, and hydrogen particle
density, $n_{H}$. This is assuming a fixed helium abundance of 10\% (by
number), a temperature of $10^{4}$ K, no turbulent velocity, and
$n(\metT)/n_{He^{+}}=10^{-7}$. Using a typical turbulent velocity of
$\xi=40~{\rm km~s^{-1}}$ would reduce the optical depth by a factor of 6.2.
The scaling of optical depth with these quantities is given in equation 1
and 2.  }
\label{fig1}
\end{figure}

For a spherically symmetric nebula, the mean probability of escape of
a photon with a frequency characterized by $x=(\nu-\nu_{o})/\Delta
\nu$ is given by (Cox \& Mathews 1969)

\begin{equation}
p(\tau_{x})=\frac{3}{4\tau_{x}}
\left[1-\frac{1}{2\tau_{x}^{2}}+\left(\frac{1}{\tau_{x}}+\frac{1}{2\tau_{x}^{2}}\right)e^{-2\tau_{x}}\right]
\end{equation}

\noindent where $\nu_{o}$ is the line center frequency and $\Delta
\nu=(\nu_{0}/c)\sqrt{2kT/m_{He}+\xi^{2}}$.

Integrating this escape probability over the entire line profile
yields a total escape probability of
$\epsilon(\tau)=\frac{1}{\sqrt{\pi}}\int_{-\infty}^{\infty}
p(\tau_{x})e^{-x^{2}}dx$. As noted in Osterbrock (1989), this can be
approximated to within 10\% with the function
$\epsilon(\tau)=1.72/(1.72+\tau)$. To get a more exact result, we have
evaluated this integral numerically, with a precision of better than 0.01\%. 

The calculation of statistical equilibrium given in equation (4) of
Paper I is then simply modified by replacing the spontaneous decay
rates $A$ with $\epsilon(\tau)A$, yielding a revised set of level
populations and emissivities. In the calculations here, we used a
model atom with individual levels up to $n=20$; otherwise the
calculations are the same as in Paper I. The relative optical
depths of lines in the $n~^{3}P$-\metT series are given in Table 1.

\section{Results}

\subsection{Tables, Fitting Formulae and Comparisons}

We have calculated the emissivity, $j=j$ ($n, T, \tau_{3889}$), of
recombination lines for the same temperature and density grid as in
Paper I, but for optical depths up to $\tau_{3889}=100$. The ratio of
the line emissivity for optical depth $\tau_{3889}$ to the emissivity
for zero optical depth is given by optical depth factor
$f_{line}(\tau_{3889})= j(n, T, \tau_{3889})/j(n, T, 0)$. Figure 2 shows how the
optical depth factor of several lines varies as a function of optical
depth for a case with $n_{e}=10^{2}~{\rm cm^{-3}}$ and $T=10000$ K and
20000 K. Although we include the collisional coupling between the
triplet and singlet levels, only the triplet lines show noticeable
differences.  Even for densities as high as $n_{e}=10^{8}~{\rm
cm^{-3}}$, the effect of optical depth in the \metT level on the
singlet lines is small, less than 0.4\% in the worst case.

\begin{figure}[ht!]
\includegraphics[angle=0,totalheight=\smpicsize]{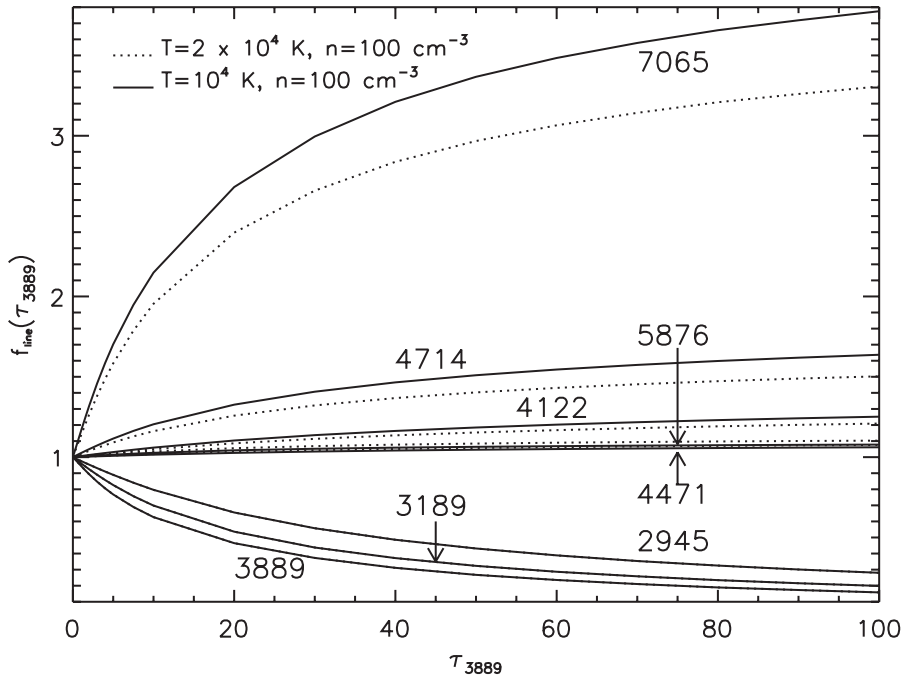}
\caption{Optical depth correction factor for selected optical lines for
cases with a particle density of $n_{H}=100~{\rm cm^{-3}~}$ and two
different temperatures.   }
\label{fig2}
\end{figure}

Table 2 shows all optical and IR lines that change by more than 5\%
over the range $\tau_{3889}=0-100$ for the case with
$n_{e}=10^{2}~{\rm cm^{-3}}$ and $T=10000$ K. This provides a guide to
which lines are most affected by optical depth.  From Figure 2, note 
the strong dependence of the 7065 \AA\  line on both $\tau_{3889}$ 
and temperature.
The decrease in $f_{7065}(\tau_{3889})$ with increasing temperature (at constant
density) arises due to a decreasing population in the \metT level
caused by the increasing collisional excitations out of that level.
We also provide
polynomial fit coefficients in Table 3 for these lines that
approximate the calculated results to within 10\%, and better than 1\%
for a few selected optical lines at a single temperature and density. 

Unfortunately, the optical depth factor is sometimes a complex
function of density, temperature, and $\tau_{3889}$, so no simple
functional form was found that could provide a good (i.e., better than
10\%) fit for $f_{line}(\tau_{3889})$ over our full three dimensional parameter
space. This difficulty is illustrated by Figure 3 which shows
$f_{7065}$ at $\tau_{3889}=100$. As a result, we have written a program
which does a linear interpolation on our grid of results to
estimate the emissivities. This program may be obtained from the
authors.\footnote{Program and input data file may be obtained from 
the website http://wisp.physics.wisc.edu/$\sim$benjamin.}

\begin{figure}[ht!]
\includegraphics[angle=0,totalheight=\smpicsize]{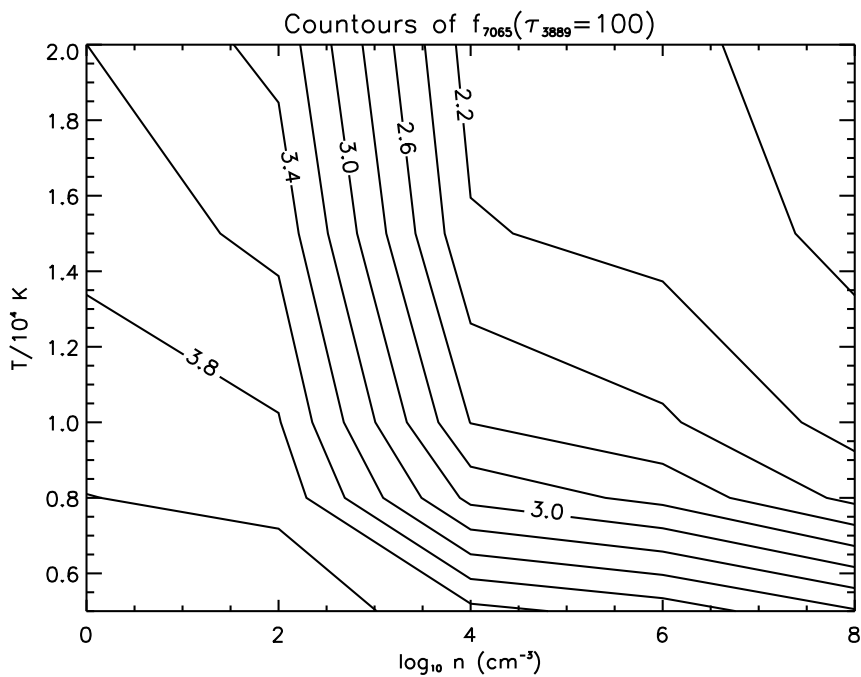}
\caption{Value of optical depth correction factor for the 7065 \AA\  line
at $\tau_{3889}=100$ for a range of temperatures and densities. There is a
large range of variation that is not well fit by a simple function. As a
result, those desiring accurate estimates of these lines should use the
FORTRAN program available from the authors.  }
\label{fig3}
\end{figure}

In Figure 4, we have compared the results of our new calculations with
the results of Robbins (1968) for a density of $n_e=100~{\rm
cm^{-3}}$.  Robbins' results are shown as filled symbols, ours are
shown as open symbols, and a solid line shows the result of the
fitting formulae in Table 3.  This figure shows differences
between the relative line strengths for different temperatures,
particularly for $\lambda$7065.  This demonstrates again that the
radiative transfer effects are dependent on other factors beside
$\tau_{3889}$. As a result, the fitting formulae given in Table 3
should be used with caution, since they are only for a single choice
of temperature and density.

\begin{figure}[ht!]
\includegraphics[angle=0,totalheight=\bigpicsize]{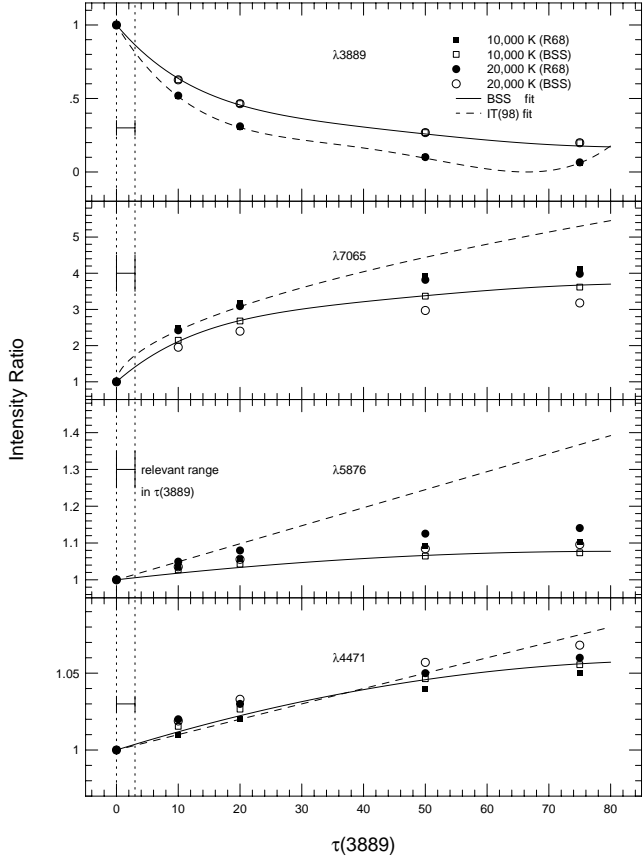}
\caption{ Comparison of recent work to earlier work of Robbins (1968) and
the Izotov \& Thuan (1998) fits to the Robbins (1968) calculations.  The
dotted vertical lines indicate the range of values of $\tau$(3889) that are
important for the giant extragalactic HII regions that are used for
determining the primordial helium abundance.  Note that the range in
intensity ratio on the y-axis changes signficantly from panel to panel.
In some cases (e.g., $\lambda$3889) the differences in the results for
the two different temperatures are small compared to the y-axis range,
and thus, the circular and square symbols are over-plotted. }
\label{fig4}
\end{figure}

Second, although Robbins only reported results on a very coarse grid,
it is instructive to note the difference in $f_{3889}$ as a function
of $\tau_{3889}$.  Our calculated $\lambda$3889 line strengths do not
decrease as strongly with increasing $\tau_{3889}$.  Since
$\lambda$3889 is essentially a resonance line, most $\lambda$3889
absorptions are followed directly by a $\lambda$3889 emission, with
the result that the line strength falls off relatively slowly with
increasing $\tau_{3889}$ (less than 50\% at $\tau_{3889}$ $=$ 10).
The absolute value of $f_{3889}$ as a function of $\tau_{3889}$, therefore, 
depends on the formalism used in the calculation.  Robbins (1968) used
an ``exact redistribution'' function from Unno (1952), while here we
use complete redistribution.  In the complete redistribution case, the
absorbed wavelength and re-emitted wavelength are completely
unrelated.  In addition, Robbins (1968) used a numerical integration
of the equations of radiative transfer while our result is (almost)
analytical, and therefore has much higher numerical accuracy.  Thus,
we believe our calculations should be the more accurate.

A significant advantage to using an escape probability formalism as opposed to
calculating the ``exact'' solution as in Robbins (1968) is that it allows
one to quickly change the characteristics of the radiative transfer.
Since these effects depend upon the exact geometry of the
system, which will always be poorly known, our ability to correct line
fluxes for modifications due to self-absorption will always be limited.
In particular, density inhomogeneities, ionization inhomogeneities, large
scale velocity structure, and planar versus spherical geometry can all
effect the resulting radiative transfer to differing degrees. Our
calculations also do not include the effects of dust absorption or the
effects of the hydrogen emission line at 3889.1 \AA.

In lieu of calculating a large number of grids of differing geometries
and density/velocity structure, we have attempted to estimate the
robustness of our results using two alternate forms of the escape
probability corresponding to differing geometries and scattering
assumptions.  For the case with $T=10^{4}$ K and $n_{e}=10^{-2}~{\rm
cm^{-3}}$, we have recalculated the recombination cascades using the
simple form $\epsilon=(1-\exp(-\tau))/\tau$ for a plane parallel
geometry(Rybicki 1984; Almog \& Netzer 1989); the second form comes
from integrating the local escape probability given by equation I.2 in
Avrett \& Hummer (1965). We find that for the restricted range of
optical depth of interest for primordial helium abundance studies, the
difference in all the bright optical lines is less than 1\%. For the
extended range to $\tau_{3889}=100$, all the triplet lines that are
not part of the $2 ^{3}S-n^{3}P$ series are modified by 7\% at the
most. Lines in the $2^{3}S-n^{3}P$ series are affected much more. Like
other lines, they are modified indirectly by the change in the $2
^{3}S$ level population (which is a small fractional change in the
line flux) , but also directly by the reprocessing of photons in this
series.

Another important effect for radiative transfer is the presence or
absence of velocity gradients. Our treatment implicitly assumes that
the expansion rate of a nebula is significantly less than the
intrinsic Doppler width of the line, i.e., $w=v_{exp}/v_{dopp}<<1$,
where $v_{dopp}=\sqrt{166T_{4}/m_{He}+\zeta^{2}}$ as above. For
extragalactic HII regions, we are generally justified in assuming that
$w \cong 0$. However, using the results of Robbins (1968), one can estimate the effects of expansion
velocity for an uniform density sphere. We find that the effects of expansion can be approximated by replacing $\tau_{3889}$ with $\tau_{3889,eff} \cong
\tau_{3889}/(1+0.4w)$. This approximation is good to within 5\%
up to $\tau_{3889}=30$ and 20\% for $\tau_{3889}=70$.

\subsection{Relevance to Primordial Helium Determinations}

Part of the motivation of this work was to estimate accurate radiative
transfer corrections in determinations of helium abundances in low
metallicity extragalactic HII regions.  Thus, as in Paper I, we have
calculated optical depth correction factors over a limited grid of
physical parameters: $T=12,000-20,000$ K, $n_{e}=1-300~ {\rm
cm^{-3}~}$, and $\tau_{3889}<$ 2.0.  The upper bound on the optical
depth is chosen for two reasons.  First, most observations of
extragalactic HII regions find optical depths less than 2.0 (e.g.,
Izotov \& Thuan 1998). This is due to the large ($\ge$ 25 km s$^{-1}$)
velocity dispersion typical of these objects (e.g., Melnick et al.\
1988, and references therein).  Second, if an object requires a large
optical depth correction to derive the helium abundance, it is
probably an inherently uncertain abundance calculation, and may not
therefore be useful in determining the primordial helium abundance.

The upper bound on density follows from the same principle, i.e., as
the correction for collisional excitation becomes large, the
uncertainty in that correction must also grow, rendering the abundance
determination less certain.  While the density range covers the entire
range of densities considered for NGC 346 by Peimbert et al.\ (2000)
and almost all of the targets studied by Izotov \& Thuan (1998),
Peimbert, Peimbert, \& Luridiana (2001) have derived much higher
densities for several of the targets in the Izotov \& Thuan (1998)
sample.  If it proves necessary to consider a grid with a larger
density range, it is possible to produce one with the provided
programs.

We find that the optical depth correction factors for the bright optical
lines 4471 \AA\  and 5876 \AA\  are less than 1\% in this limited range. 
It is the corrections to 7065 \AA\  and 3889 \AA\  that
are of most concern. In Table 4, we give fitting functions that allow one to 
estimate the helium abundance given an observed line ratio of $r_{line}=
I_{line}/I_{H \beta}$. The helium abundance by number for a given line is given by 
\begin{equation}
y_{line}=r_{line} \frac{F_{line}(n_{e},t)}{f_{line}(n_{e},t,\tau_{3889})}
\end{equation}  
\noindent The function $F_{line}(n_{e},t)=At^{B_{0}+B_{1}n_{e}}$ is taken from Paper I, while the optical depth 
function,  $f_{line}(n_{e},t,\tau_{3889})=1+(\tau/2)[a+(b_{0}+b_{1}n_{e}+b_{2}n_{e}^{2})t]$ 
is calculated here for the limited range $\tau \leq 2.0$. For the 7065 \AA\ line
however, it is necessary to use a different functional form. The 
$H \beta$ emissivity is fit by 
$\epsilon_{H \beta}(n_{e},t)=4 \pi j_{H \beta}/n_{e}n_{H^{+}}=12.450 \times 10^{-26} t^{-0.917}$ with an maximum error of 
0.3\% over our limited parameter space. The 7065 emissivity at $\tau=0$ 
is fit by $\epsilon_{7065}(n_{e},t)=4 \pi j_{7065}/n_{e}n_{He^{+}}= C+(D_{0}+D_{1}n_{e}+D_{2}n_{e}^{2})t$, where 
$(C,D_{0},D_{1},D_{2})=(3.4940, -0.793, 1.50 \times 10^{-3}, -6.96 \times 10^{-7})\times 10^{-26}$, 
and the optical depth function $f_{7065}(n_{e},t,\tau_{3889})$ is the same as above with parameters $(a,b_{0},b_{1},b_{2})=(0.359, -3.46 \times 10^{-2}, -1.84 \times 10^{-4}, 3.039 \times 10^{-7})$. Then 
the helium abundance by number using the 7065 line is given by

\begin{equation}
y_{7065}=r_{7065}\frac{\epsilon_{H \beta}(n_{e},t)}{\epsilon_{7065}(n_{e},t)f_{7065}(n_{e},t,\tau_{3889})}
\end{equation}

\noindent These functions fit $y_{7065}$ within 2.2\% over our limited range.   


Figure 5 shows the same functions as in Figure 4 but only for the low
values of $\tau_{3889}$ relevant to estimating the primordial helium
abundance. We include the fits to the optical depth correction factors contained
in Table 4, and refer to them as ``BSS limited fit''. A striking feature in Figure 5 is the large deviation of
the Izotov \& Thuan fitting formula from our newly calculated values
for the 7065 \AA\ line.  Since the Izotov \& Thuan fit is based on the
very coarse grid of values given by Robbins, it does not accurately
represent the results at low $\tau_{3889}$ where it is actually being
applied.  The parabolic form of the fitting function is not
appropriate to the physics, as $f$ increases nearly linearly with
$\tau_{3889}$ for low values of $\tau_{3889}$, and turns over only at
relatively high values of $\tau_{3889}$.

\begin{figure}[ht!]
\includegraphics[angle=0,totalheight=\bigpicsize]{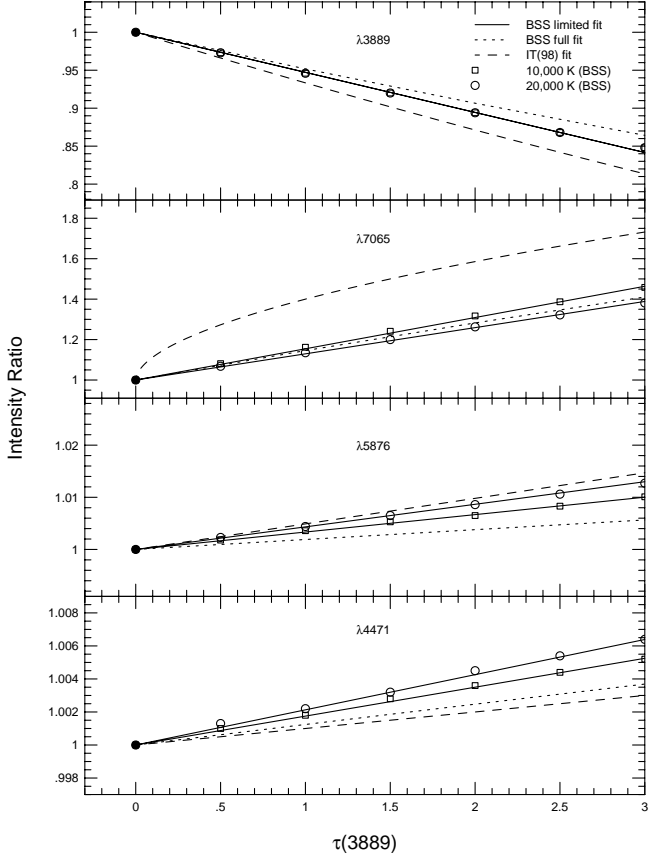}
\caption{A plot similar to that shown in Figure 4, zooming in on the
range of $\tau$(3889) from 0 to 3.  Note that the y-axis ranges are different
from those of Figure 4, and, again, change from panel to panel. 
The striking feature in this figure is the large difference between
the new calculated values for $\lambda$7065 and the interpolation
fit to the Robbins (1968) calculations derived by Izotov \& Thuan (1998).
For low values of $\tau$(3889), the new calculations show that the
amplification of the $\lambda$7065 line is much lower than would have been
inferred from the Izotov \& Thuan (1998) fit.  This implies that the
$\lambda$7065 line is a better density diagnostic than previously
thought.  As in Figure 4, the differences in the results for
the two different temperatures for $\lambda$3889 are small compared to 
the y-axis range, and thus, the circular and square symbols are over-plotted. 
}
\label{fig5}
\end{figure}

Evidence of this can be seen by considering the predictions of the
Izotov \& Thuan functions at $\tau_{3889}=0.5$.  Figure 5 shows that at
$\tau_{3889} = 0.5$ the 3889 \AA\ line has decreased by roughly 3\%
while the Izotov \& Thuan fit for the 7065 \AA\ line has increased by
roughly 25\%.  At 20,000 K, the ratio of the emissivities for the
$\lambda$3889 and $\lambda$7065 lines is 3.85, which, after correcting
for energy per photon, means that there are roughly 2.1 $\lambda$3889
photons emitted for each $\lambda$7065 photon.  Since the increase in
the 7065 \AA\ line with increasing $\tau_{3889}$ comes from the
conversion of $\lambda$3889 photons, the 7065 \AA\ line should not
increase faster than 2.1 $\times$ the 3889 \AA\ line decreases.  Thus,
in the range of interest, the Izotov \& Thuan fit has overestimated
the radiative transfer effects on the 7065 \AA\ line by roughly a
factor of four at $\tau_{3889} = 0.5$, and higher at lower values of
$\tau_{3889}$.

This difference has important consequences.  Olive \& Skillman (2001)
explored various possibilities for systematic errors that could result
by using minimization routines to solve for physical conditions and He
abundances simultaneously.  Since the 7065 \AA\ line plays the main
role as the density diagnostic in these minimization routines,
overestimating the 7065 \AA\ dependence on $\tau_{3889}$ leads to
errors in the density estimates (which directly affect the helium
abundance determinations).  The good news is that the 7065 \AA\ line
dependence on $\tau_{3889}$ is much less than previously thought, so
that using the 7065 \AA\ as a density diagnostic will increase in
diagnostic power.  That is, the degeneracies between the 3889 \AA\ and
7065 \AA\ lines will be decreased.  This is important since the 3889
\AA\ line is subject to systematic observational uncertainties, most
notably separation from the 3889 HI line.

The error estimation work of Olive \& Skillman (2001) will need to be
revisited with the newly calculated fitting formulae.  We recommend
that all future helium abundance calculations from minimization
routines take advantage of the new fitting formulae presented in Table
4.  The consequence of our results on the determination of the
primordial helium abundance is not yet clear.  Only about one third of
the targets in the sample of Izotov \& Thuan (1998) showed evidence
for detectable radiative transfer effects.  Moreover, since the main
effect comes through the densities calculated from the 7065 \AA\ line,
the effect is likely to be small in most cases. The total effect will
need to be calculated on a case-by-case basis.  However, the sense of
the correction will be that higher 7065 \AA\ line strengths will be
interpreted as being due to higher densities, not higher values of
$\tau_{3889}$. As can be seen using the helium abundance fitting
functions in Table 9 of Paper I, this will result in a lower helium
abundance for objects for which radiative transfer corrections have
been applied.

\acknowledgments We would like to thank John Mathis for comments and
encouragement, and Manuel Peimbert and Stuart Pottasch for pointing
out errors in Paper I.  We would also like to acknowledge the support
of NASA Astrophysics Theory Grant NAG5-8417 (RAB) and NASA LTSARP
grant NAG5-9221 and the University of Minnesota (EDS).

\appendix
\section{Corrections to Paper I}

Two corrections to the fitting formulae provided in Paper I should be
noted. First, the fitting formula for primordial helium abundance in
Table 9 for the 6678 line should be given by $f=2.58 t^{0.249-n_{e}2.0
\times 10^{-4}}$. Second, the values of fluxes in Tables 5-7 for
Brackett $\gamma$ emissivities were in error. The values of the
emissivities in units of $10^{-27}~{\rm ergs~s^{-1}~cm^{3}}$ for
$T=(5000,10000,20000~{\rm K})$ should be (7.25,3.44,1.54),
(7.20,3.41,1.53), and (6.99,3.33,1.51) for $n_{e}=10^{2},~10^{4}, {\rm
and~} 10^{6}~{\rm cm^{-3}}$.  The corresponding fitting functions
(good to within 8\%) are $3.39 \times 10^{-27} t^{-1.118}$, $3.37
\times 10^{-27} t^{-1.116}$, and $3.29 \times 10^{-27} t^{-1.104}$.

\clearpage
\begin{deluxetable}{rrrrrr}
\tablecolumns{6}
\tablewidth{0pc}
\tablecaption{Optical depth ratios of 2 $^{3}S$-n $^{3}P$ series\tablenotemark{a}}
\tablehead{
$n_{u}$ & $\lambda$(\AA) & $\tau$ & 
$n_{u}$ & $\lambda$(\AA) & $\tau$ 
}
\startdata
   2  &  10833   &   2328.35 &  12  &   2654    &     0.86    \\
   3  &   3890   &    100.00 &  13  &   2646    &     0.66    \\
   4  &   3189   &     32.77 &  14  &   2639    &     0.51    \\ 
   5  &   2946   &     14.68 &  15  &   2634    &     0.42    \\
   6  &   2830   &      7.88 &  16  &   2630    &     0.34    \\
   7  &   2765   &      4.74 &  17  &   2627    &     0.30    \\
   8  &   2724   &      3.08 &  18  &   2624    &     0.23    \\
   9  &   2697   &      2.12 &  19  &   2621    &     0.21    \\
  10  &   2678   &      1.53 &  20  &   2619    &     0.18    \\
  11  &   2664   &      1.10 & \nodata	    &	\nodata	& \nodata \\
\enddata
\tablenotetext{a}{Normalized to $\tau_{3890}=100$.}
\label{tbl-1}
\end{deluxetable}

\clearpage
\begin{deluxetable}{rrrcccc}
\tabletypesize{\scriptsize}
\tablecolumns{7}
\tablewidth{0pc}
\tablecaption{Optical Depth Correction $f_{line}(\tau_{3889})$ for He I Emission Line Intensities \tablenotemark{a}}

\tablehead{ \colhead{$\lambda$ (\AA)}     &\colhead{$n_{u}$} & \colhead{$l_{u}$}      
& \colhead{$\tau=2$}   
& \colhead{$\tau=10$}
& \colhead{$\tau=60$}           
& \colhead{$\tau=100$} 
}
\startdata
    2946 &   5 &  1 &    0.96 &    0.81 &    0.40 &    0.29 \\
    3189 &   4 &  1 &    0.93 &    0.71 &    0.29 &    0.19 \\
    3733 &   7 &  0 &    1.00 &    1.01 &    1.04 &    1.06 \\
    3868 &   6 &  0 &    1.00 &    1.02 &    1.09 &    1.12 \\
    3890 &   3 &  1 &    0.90 &    0.64 &    0.22 &    0.14 \\
    4122 &   5 &  0 &    1.01 &    1.06 &    1.20 &    1.25 \\
    4472 &   4 &  2 &    1.00 &    1.01 &    1.05 &    1.06 \\
    4714 &   4 &  0 &    1.04 &    1.19 &    1.54 &    1.64 \\
    5877 &   3 &  2 &    1.01 &    1.03 &    1.07 &    1.08 \\
    7067 &   3 &  0 &    1.30 &    2.10 &    3.52 &    3.81 \\
    7162 &  10 &  1 &    1.00 &    1.04 &    1.24 &    1.37 \\
    7300 &   9 &  1 &    1.00 &    1.06 &    1.32 &    1.47 \\
    7501 &   8 &  1 &    1.01 &    1.09 &    1.45 &    1.63 \\
    7818 &   7 &  1 &    1.02 &    1.14 &    1.63 &    1.84 \\
    8363 &   6 &  1 &    1.04 &    1.24 &    1.91 &    2.16 \\
    9012 &  10 &  1 &    1.00 &    1.04 &    1.24 &    1.37 \\
    9231 &   9 &  1 &    1.00 &    1.06 &    1.32 &    1.47 \\
    9466 &   5 &  1 &    1.10 &    1.46 &    2.41 &    2.69 \\
    9555 &   8 &  1 &    1.01 &    1.09 &    1.45 &    1.63 \\
    9705 &   7 &  0 &    1.00 &    1.01 &    1.04 &    1.06 \\
   10075 &   7 &  1 &    1.02 &    1.14 &    1.63 &    1.84 \\
   10671 &   6 &  0 &    1.00 &    1.02 &    1.09 &    1.12 \\
   10833 &   2 &  1 &    1.02 &    1.07 &    1.17 &    1.19 \\
   10999 &   6 &  1 &    1.04 &    1.24 &    1.91 &    2.16 \\
   12531 &   4 &  1 &    1.26 &    2.03 &    3.56 &    3.93 \\
   12850 &   5 &  0 &    1.01 &    1.06 &    1.20 &    1.25 \\
   12988 &   5 &  1 &    1.10 &    1.46 &    2.41 &    2.69 \\
   14492 &  10 &  1 &    1.00 &    1.04 &    1.24 &    1.37 \\
   15067 &   9 &  1 &    1.00 &    1.06 &    1.32 &    1.47 \\
   15951 &   8 &  1 &    1.01 &    1.09 &    1.45 &    1.63 \\
   17006 &   4 &  2 &    1.00 &    1.01 &    1.05 &    1.06 \\
   17382 &  10 &  1 &    1.00 &    1.04 &    1.24 &    1.37 \\
   17455 &   7 &  1 &    1.02 &    1.14 &    1.63 &    1.84 \\
   18214 &   9 &  1 &    1.00 &    1.06 &    1.32 &    1.47 \\
   19523 &   8 &  1 &    1.01 &    1.09 &    1.45 &    1.63 \\
   19550 &   4 &  1 &    1.26 &    2.03 &    3.56 &    3.93 \\
   20428 &   6 &  1 &    1.04 &    1.24 &    1.91 &    2.16 \\
   21128 &   4 &  0 &    1.04 &    1.19 &    1.54 &    1.64 \\
\enddata
\tablenotetext{a}{These correction factors for case with $n_{e}=10^{2}~cm^{-3}$ and $T=10^{4}$ K. 
Only lines from $n_{u} \leq 10$ with  $1.05 \geq f(\tau_{3889}) \geq 0.95$ over the range 
$\tau_{3889}=[0,100]$ are included. }
\label{tbl-2}
\end{deluxetable}

\clearpage
\begin{deluxetable}{rrrrrrrr}
\tabletypesize{\scriptsize}
\tablecolumns{8}
\tablewidth{0pc}
\tablecaption{Fitting function for $f_{line}(\tau_{3889})$ for $n_{e}=10^{2}~{\rm cm^{-3}}$ and $T=10^{4}$ K.  \tablenotemark{a}}

\tablehead{ \colhead{$\lambda$ (\AA)}     &\colhead{$f_{0}$} & \colhead{$10^{2}f_{1}$}      
& \colhead{$10^{4}f_{2}$}   
& \colhead{$10^{6}f_{3}$}
& \colhead{$10^{8}f_{4}$}           
& \colhead{$10^{10}f_{5}$} 
& \colhead{Max Err (\%)}
}
\startdata
    2946 &    1.000 &   -2.054 &    2.398 &   -1.066 &    \nodata &    \nodata &          3.2 \\
    3189 &    1.000 &   -3.479 &    7.466 &   -7.871 &    3.082 &    \nodata &            4.8 \\
    3733 &    1.005 &    \nodata &    \nodata &    \nodata &    \nodata &    \nodata &    5.4 \\
    3868 &    1.010 &    \nodata &    \nodata &    \nodata &    \nodata &    \nodata &    9.6 \\
    3890 &    1.000 &   -4.983 &   16.086 &  -28.586 &   25.038 &   -8.415 &              5.0 \\
    4122 &    1.003 &    0.288 &    \nodata &    \nodata &    \nodata &    \nodata &      3.7 \\
    4472 &    1.000 &    0.125 &   -0.067 &    \nodata &    \nodata &    \nodata &        0.3 \\
    4714 &    1.003 &    1.485 &   -0.896 &    \nodata &    \nodata &    \nodata &        4.3 \\
    5877 &    1.000 &    0.192 &   -0.119 &    \nodata &    \nodata &    \nodata &        0.7 \\
    7067 &    1.000 &   15.008 &  -46.190 &   80.374 &  -69.603 &  -23.231 &              1.3 \\
    7162 &    0.999 &    0.385 &    \nodata &    \nodata &    \nodata &    \nodata &      1.4 \\
    7300 &    1.000 &    0.509 &    \nodata &    \nodata &    \nodata &    \nodata &      2.3 \\
    7501 &    1.002 &    0.687 &    \nodata &    \nodata &    \nodata &    \nodata &      3.9 \\
    7818 &    1.005 &    0.947 &    \nodata &    \nodata &    \nodata &    \nodata &      6.8 \\
    8363 &    1.001 &    2.231 &   -1.111 &    \nodata &    \nodata &    \nodata &        2.7 \\
    9012 &    0.999 &    0.385 &    \nodata &    \nodata &    \nodata &    \nodata &      1.4 \\
    9231 &    1.000 &    0.509 &    \nodata &    \nodata &    \nodata &    \nodata &      2.3 \\
    9466 &    1.006 &    3.778 &   -2.187 &    \nodata &    \nodata &    \nodata &        6.8 \\
    9555 &    1.002 &    0.687 &    \nodata &    \nodata &    \nodata &    \nodata &      3.9 \\
    9705 &    1.005 &    \nodata &    \nodata &    \nodata &    \nodata &    \nodata &    5.4 \\
   10075 &    1.005 &    0.947 &    \nodata &    \nodata &    \nodata &    \nodata &      6.8 \\
   10671 &    1.010 &    \nodata &    \nodata &    \nodata &    \nodata &    \nodata &    9.6 \\
   10833 &    1.005 &    0.242 &    \nodata &    \nodata &    \nodata &    \nodata &      5.2 \\
   10999 &    1.001 &    2.231 &   -1.111 &    \nodata &    \nodata &    \nodata &        2.7 \\
   12531 &    1.007 &   10.565 &  -14.802 &    7.238 &    \nodata &    \nodata &          6.0 \\
   12850 &    1.003 &    0.288 &    \nodata &    \nodata &    \nodata &    \nodata &      3.7 \\
   12988 &    1.006 &    3.778 &   -2.187 &    \nodata &    \nodata &    \nodata &        6.8 \\
   14492 &    0.999 &    0.385 &    \nodata &    \nodata &    \nodata &    \nodata &      1.4 \\
   15067 &    1.000 &    0.509 &    \nodata &    \nodata &    \nodata &    \nodata &      2.3 \\
   15951 &    1.002 &    0.687 &    \nodata &    \nodata &    \nodata &    \nodata &      3.9 \\
   17006 &    1.006 &    \nodata &    \nodata &    \nodata &    \nodata &    \nodata &    5.1 \\
   17382 &    0.999 &    0.385 &    \nodata &    \nodata &    \nodata &    \nodata &      1.4 \\
   17455 &    1.005 &    0.947 &    \nodata &    \nodata &    \nodata &    \nodata &      6.8 \\
   18214 &    1.000 &    0.509 &    \nodata &    \nodata &    \nodata &    \nodata &      2.3 \\
   19523 &    1.002 &    0.687 &    \nodata &    \nodata &    \nodata &    \nodata &      3.9 \\
   19550 &    1.007 &   10.565 &  -14.802 &    7.238 &    \nodata &    \nodata &          6.0 \\
   20428 &    1.001 &    2.231 &   -1.111 &    \nodata &    \nodata &    \nodata &        2.7 \\
   21128 &    1.003 &    1.485 &   -0.896 &    \nodata &    \nodata &    \nodata &        4.3 \\
\enddata
\tablenotetext{a}{$f(\tau_{3889})=f_{0}+f_{1}\tau+f_{2}\tau^{2}...$.  
Only lines from $n_{u} \leq 10$ with  $1.05 \geq f(\tau_{3889}) \geq 0.95$ over the range 
$\tau_{3889}=[0,100]$ are included. }
\label{tbl-3}
\end{deluxetable}

\clearpage
\begin{deluxetable}{rrrrrrrrr}
\tabletypesize{\scriptsize} 

\tablecolumns{9}
\tablewidth{0pc}
\tablecaption{Limited Range Fitting Formulae For Helium Abundance Including Optical Depth Effects  \tablenotemark{a}}

\tablehead{ 

\colhead{Line}     &\colhead{A}  
&\colhead{$B_0$} &\colhead{$B_1$}      
& \colhead{$a$}   
& \colhead{$b_0$}
& \colhead{$b_1$}
& \colhead{$b_2$}
& \colhead{Max Error} \\
\colhead{(\AA)}     &\colhead{}  
&\colhead{} &\colhead{}      
& \colhead{}   
& \colhead{}
& \colhead{}
& \colhead{}
& \colhead{(\%)}
}
\startdata
    3889 & 0.904  & -0.173 & $-5.4 \times 10^{-4}$  &$-1.06 \times 10^{-1}$ & $5.14 \times 10^{-5}$  & $-4.20\times 10^{-7}$  & $1.97 \times 10^{-10}$   &  0.8\% \\
    4026 & 4.297   & 0.090  & $-6.3 \times 10^{-6}$  & $1.43 \times 10^{-3}$ & $4.05 \times 10^{-4}$  & $3.63 \times 10^{-8}$ &    \nodata                    &     0.2\% \\
    4387 & 16.255   & 0.109 & $-3.7 \times 10^{-6}$  &     \nodata                   &     \nodata                    &  \nodata                       &  \nodata                      &     0.2\% \\
    4471 & 2.010   & 0.127  & $-4.1 \times 10^{-4}$  & $2.74 \times 10^{-3}$ & $8.81 \times 10^{-4}$  & $-1.21 \times 10^{-6}$ &    \nodata                    &     1.0\% \\
    4686 & 0.0816 & 0.145  &   \nodata                     &    \nodata                    &      \nodata                   &         \nodata                &    \nodata                    &     0.3\% \\
    4714 & 19.555 & -0.461 & $-13.5 \times 10^{-4}$ & $5.77 \times 10^{-2}$ & $-7.12 \times 10^{-3}$ & $-3.10 \times 10^{-5}$ & $3.99 \times 10^{-8}$ &     1.9\% \\
    4922 & 7.523   & 0.150  & $-1.9 \times 10^{-4}$  &     \nodata                   &     \nodata                    &  \nodata                       &  \nodata                      &     0.6\% \\
    5876 & 0.735  & 0.230  & $-6.3 \times 10^{-4}$  & $4.70 \times 10^{-3}$ & $2.23 \times 10^{-3}$  & $-2.51 \times 10^{-6}$ &    \nodata                    &     1.2\% \\
    6678 & 2.580   & 0.249  & $-2.0 \times 10^{-4}$  &     \nodata                   &     \nodata                    &  \nodata                       &  \nodata                      &     0.3\% \\
    7065 & See text  &  &  &  &  &  &     &   \\
    7281 & 13.835   & -0.307  & $-7.0 \times 10^{-4}$  &     \nodata                   &     \nodata                    &  \nodata                       &  \nodata                      &     2.0\% \\
\enddata
\tablenotetext{a}{Helium number abundance is given by $y=(I_{line}/I_{H\beta}) F(n_{e},T)/ f(n_{e},T,\tau_{3889})$, where $I_{line}/I_{H\beta}$ is the observed line ratio, $F(n_{e},T)=At^{B_{0}+B_{1}n_{e}}$ (from Paper I), and the optical depth correction factor is given by $f(n_{e},T,\tau_{3889})=1+(\tau_{3889}/2)[a+(b_0+b_1n_e+b_2n_{e}^{2})t]$. These functions apply \underline{only} for cases in the range $n_{e}=1- 300~{\rm cm^{-3}}$ and $T=1.2-2.0 \times 10^{4}$ K. The maximum uncertainty over
this range is noted.}
\label{tbl-4}
\end{deluxetable}


\end{document}